\documentclass[aps,pre,amsmath,twocolumn,amssymb,floatfixng,showpacs,longbibliography,
superscriptaddress,nofootinbib]{revtex4-2}
\usepackage{amsfonts}
\usepackage{amsmath}
\usepackage{amssymb}
\usepackage{latexsym}
\usepackage{array}
\newcolumntype{P}[1]{>{\centering\arraybackslash}p{#1}}
\newcolumntype{M}[1]{>{\centering\arraybackslash}m{#1}}
\usepackage{caption}
\usepackage[colorlinks,citecolor=red,urlcolor=blue,bookmarks=false,hypertexnames=true]{hyperref} 
\usepackage{xcolor}
\usepackage{bm}
\usepackage{relsize}
\usepackage{float}
\usepackage{subfloat}
\usepackage{lipsum}
\usepackage{dcolumn}
\usepackage{graphicx}
\usepackage{multirow}
\usepackage{makecell}
\usepackage{mathtools}
\usepackage[bbgreekl]{mathbbol}

\usepackage{footnote}

\begin{document}

\title{Quantum Langevin dynamics and the long-time behaviour of two charged coupled oscillators in a common heat bath}
\author{Koushik Mandal}
\affiliation{Department of Applied Science, Haldia Institute of Technology, Haldia, 721657, India}

\author{Suraka Bhattacharjee}
\email{surakabhattacharjee@eee.sastra.edu}
\affiliation{$^{,a}$Department of Physics, School of EEE, SASTRA Deemed University, Tirumalaisamudram,
Thanjavur - 613401, Tamilnadu, India \\
$^b$Department of Theoretical Physics, Raman Research Institute, Sadashivanagar, Bangalore-560080, Karnataka, India}           


\date{\today}

\begin{abstract}
In this paper, the moderately long-time behaviours of the correlation functions for two charged coupled harmonic oscillators connected to a common heat bath are analyzed in the presence of a magnetic field via the Quantum Langevin dynamics. Interestingly it is seen that at long times the correlation functions at $T\rightarrow 0$ exhibit a power law decay with the coefficients of the power laws being completely different for the two masses, affecting the overall dynamics of the coupled system. The effect of the bath-induced force on mass $m_1$ mediated by the interaction of $m_2$ with the common heat bath is studied and the results are highlighted in the presence of an external magnetic field. It is shown that the effect of cyclotron frequency increases the correlation functions at an instant of time, lowering the rate of temporal decay of the correlation functions. The results in the absence of a magnetic field are also presented, which are extremely important for investigating the movements of the atoms in protein molecules at low temperatures.
\end{abstract}
\maketitle
\section{Introduction}
The correlation functions of a Brownian particle attached to a heat bath has been studied for a long time involving different anomalous bath models \cite{Alder,Kai,Felderhof,tothova2020brownian}. The Brownian motion for quantum spins coupled to heat bath has also been investigated extensively in the recent times \cite{ma2011langevin,anders2022quantum,surakaspin}. The effect of environment on a particle has been extensively studied in different aspects, including the recent study of a particle in a double well potential \cite{das2020quantum,Maile}. Further, the quantum Brownian motion has been analyzed in terms of the response function, decoherence, position autocorrelation, position-velocity correlation and velocity auto-correlation functions \cite{grabertone,graberttwo,decosup,bhattacharjee2023decoherence,bhattacharjee2022quantum}. However, it is well established that in the Classical regime, the correlation functions at long times exhibit $t^{-3/2}$  power-law decay in the bulk and a $t^{-5/2}$ decay near the boundaries, whereas, the position autocorrelation show a $t^{-2}$ decay for Quantum regime \cite{Alder, Tongcang,Franosch,Simon,Jeney,Jannasch}. Moreover, some of our recent studies are focused on the derivation of the correlation functions for Brownian motion of a charged particle using the Quantum Langevin equation coupled to an Ohmic bath, in the presence of a harmonic oscillator potential \cite{surakaone}. Our results of the position autocorrelation function showed the logarithmic decay in the absence of harmonic potential and $t^{-2}$ decay in the presence of both harmonic oscillator potential and magnetic field \cite{graberttwo,bhattacharjee2022long}. \\
\hspace*{0.2cm} In some earlier literature, the coupled oscillators had been studied in different aspects including the calculation of response function of two coupled Classical harmonic oscillators via the Langevin dynamics \cite{Berkowitz,ford1}.  It was shown that a system of large number (N) of coupled oscillators can be treated as the independent oscillator (IO) problem, where N-1 oscillators are treated as the heat bath \cite{ford1}. Some recent studies have focused on the analysis of two uncharged oscillators trapped near to each other and coupled to a common heat bath. The entanglement dynamics is investigated using the Langevin formalism \cite{fordPRA,O'Connell_2012,Zell}. In \cite{Duarte}, a similar system-plus-reservoir approach was used to study the dynamics of two Brownian particles non-linearly coupled to the reservoir. Recently the Non-equilibrium Statistical Mechanics of two trapped coupled Brownian oscillators in presence of a common heat bath was studied using the sum rule of Linear Response Theory \cite{de2014nonequilibrium}. All these results are highly effective in studying the dynamics of atoms in protein molecules, which are greatly affected by the bond stretching of the neighbouring atoms \cite{McCammon,McCammon1}. Later, the master equation and the Quantum decoherence of two coupled oscillators in the presence of a common environment were analyzed \cite{Chou,Ponte1,Ponte,Rajagopal}. The thermodynamics of two under-damped interacting  Brownian particles was also studied and the effective thermodynamic description of the first particle was given using different coarse-graining schemes like the marginalization over one particle, bipartite structure with information
flows, and the Hamiltonian of mean force formalism \cite{Herpich}.  In \cite{ford2014}, the authors have described the dynamics of two oscillators coupled to a common heat bath via the Lamb model of two oscillators connected by a string. The bath mediated-interaction between the oscillators is considered to be originated from the waves propagating in the string from one particle to the another. Nevertheless, in the present work, we have focused on the moderately long-time behaviours of position autocorrelation, position-velocity correlation and velocity autocorrelation of two charged coupled oscillators  in presence of a magnetic field and interacting with a common Ohmic heat bath of infinite number of harmonic oscillators. In the common heat bath problem the dynamics of one particle is seen to be highly affected by the interaction between the other particle and the bath \cite{de2014nonequilibrium}. We have investigated this in detail and emphasized the effect of the magnetic field on the dynamics of the coupled oscillators in the context of Quantum Brownian motion. The calculations are done at $T\rightarrow 0$ temperature i.e., the Quantum fluctuation dominated low-temperature regime. Hence, these results would be helpful in analyzing the small atoms in protein molecules at very low temperatures.\\
It can be noted that the Coulomb repulsion between the charged particles is nearly constant at the moderately
long time regimes as the variation in Coulomb repulsion
induced by the displacement of the particles  ($\Delta x (t)$ and
$\Delta y (t)$ are very small for $T\rightarrow 0$ in the diffusive regime of the Einstein’s equation $\langle 
[\Delta x]^2\rangle=2\left(K_B T/\zeta \right)t$
($\zeta$ representing the Stoke’s friction coefficient) \cite{chaikin_lubensky_1995}. So in this regime, the dynamics of the particles are primarily governed by three competing time scales of dissipation constant ($\gamma$), cyclotron frequency ($\omega_c$) and the harmonic oscillator frequency ($\omega_0$).\\
\hspace*{0.2cm} The paper is organized as follows: In Sec.II,  the Langevin equation for coupled harmonic oscillators in the magnetic field is formulated and the position autocorrelation function, position-velocity correlation and velocity autocorrelation functions are defined in the context of the Fluctuation-dissipation Theorem (FDT). In Sec.III, the calculations of the correlation functions are shown  and the magnetic field dependence are highlighted for different regimes of time. The long-time behaviours are also tabulated at the end of the section. The results are plotted and discussed in Sec. IV. The paper is concluded with an overall comment on the entire calculation and the future prospects in Sec. V. 
\section{Langevin dynamics of two coupled oscillators and a magnetic field}

\begin{figure}[H]
\centering
\includegraphics[scale=0.95]{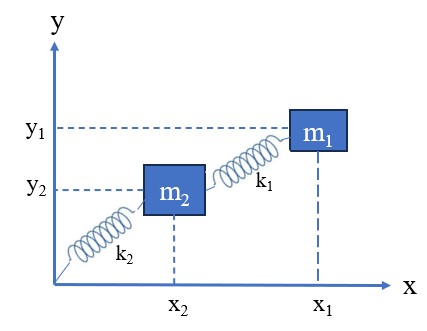} 
\caption{Two charged quantum Brownian particles coupled to each other and placed in a magnetic field while attached to a harmonic bath is schematically shown.}
\label{schematic}
\end{figure}

The Hamiltonian of two charged coupled harmonic oscillators in presence of a magnetic vector potential and connected to a heat bath via position co-ordinate coupling is given by:
\begin{align}
H=&\frac{\left( \vec{p}_1-\frac{q_1  \vec{A}_{1} (\vec{r}_1)}{c}\right)^2}{2m_1}+\frac{\left( \vec{p}_2-\frac{q_2 \vec{A}_{2}(\vec{r}_2)}{c}\right)^2}{2m_2}+ \notag \\
&\frac{1}{2}\left(k_1 \left(\vec{r}_1-\vec{r}_2\right)^2 +k_2  \vec{r}_2^2 \right)+  q_1 V(\vec{r}_1)+q_2 V(\vec{r}_2)+ \notag \\
& \sum_{j=1}^N\left[\frac{1}{2m_j} \vec{p}_{j}^2 +\frac{1}{2}m_j \omega_j^2 \left( \vec{q}_{j}-  \vec{r}_1-\vec{r}_2\right)^2 \right] \label{mainHamiltonian}
\end{align}
where $m_1$, $m_2$, $\vec{q}_1$, $\vec{q}_2$, $ \vec{r}_{1}$, $ \vec{r}_{2}$, $ \vec{p}_{1}$, $ \vec{p}_{2}$ are the masses, charges, position and momentum coordinates of the first and second Brownian particle respectively. $k_1$, $k_2$ are the spring constants of the two springs as shown in Fig.(\ref{schematic}) and  V($\vec{r}_1$), V($\vec{r}_2$)  represent the Coulomb potentials at points $\vec{r}_1$, $\vec{r}_2$ respectively. Further, $\vec{q}_j$, $\vec{p}_j$ and $\omega_j$ are the position, momentum and frequency of the $j^{th}$ bath oscillator respectively. $ \vec{A}_1(\vec{r}_1)$ and $ \vec{A}_2(\vec{r}_2)$ represent the vector potentials at  $\vec{r}_1$ and $\vec{r}_2$ and $c$ is the velocity of light. Here the last term in the Hamiltonian denotes the coupling between the position coordinate of the bath particle and the position coordinates of the Brownian particles $ \vec{r}_1$ and $ \vec{r}_2$.\\
  The coupled oscillator Hamiltonian in Eq. (\ref{mainHamiltonian}) can be explicitly written in terms of the system ($H_S$), bath ($H_B$) and interaction Hamiltonian ($H_{SB}$) \cite{PhysRevLett.100.220401}:
\begin{widetext}
\begin{align}
    &H_s=\frac{\left( \vec{p}_1-\frac{q_1  \vec{A}_{1} }{c}\right)^2}{2m_1}+\frac{\left( \vec{p}_2-\frac{q_2 \vec{A}_{2}}{c}\right)^2}{2m_2}+ \frac{1}{2}\left(k_1 \left(\vec{r}_1-\vec{r}_2\right)^2 +k_2 \vec{r}_2^2 \right)+ q_1 V\left(\vec{r}_1\right)+q_2 V\left(\vec{r}_2\right)+ \notag \\
    &\frac{1}{2}\sum_{j=1}^N m_j \omega_j^2 \vec{r}_1^2+\frac{1}{2}\sum_{j=1}^N m_j \omega_j^2 \vec{r}_2^2+ \sum_{j=1}^N m_j \omega_j^2 \left(\vec{r}_1.\vec{r}_2\right) \label{Hs}\\
   &H_B=\sum_{j=1}^N \frac{1}{2m_j} \vec{p}_j^2+\frac{1}{2}m_j \omega_j^2 \vec{q}_j^2 \label{HB}\\
   &H_{SB}= -\sum_{j=1}^N m_j \omega_j^2 \left \lbrace \vec{q}_j . (\vec{r}_1+\vec{r}_2) \right \rbrace \label{HSB}
\end{align}
\end{widetext}
The last term in the system Hamiltonian $H_s$ signifies the interaction between the two particles, modulated by the frequency of the bath oscillators.\\
Using the Eq.(\ref{mainHamiltonian}), the Heisenberg equations of motion for the particle coordinates yield:
\begin{align}
    \dot{\vec{r}}_1=\frac{1}{i\hbar}\left[\vec{r}_1, H 
 \right]=\frac{\left(\vec{p}_1-\frac{q_1 \vec{A}_1}{c}\right)}{m_1}\label{Heisenbergpaticle1}\\
  \dot{\vec{r}}_2=\frac{1}{i\hbar}\left[\vec{r}_2, H 
 \right]=\frac{\left(\vec{p}_2-\frac{q_2 \vec{A}_2}{c}\right)}{m_2}\label{Heisenbergpaticle2}
\end{align}

Similarly, using Eq.(\ref{mainHamiltonian}) for the momentum coordinates:
\begin{widetext}
\begin{align}
    \dot{\vec{p}}_1=&\frac{1}{i\hbar}\left[\vec{p}_1,H  \right]=\frac{q_1}{c}\left( \vec{v}_1\times \vec{B} \right)+\frac{q_1}{c} \left(\vec{v}_1. \vec{\nabla}_1  \right)\vec{A}_1+\frac{i \hbar q_1}{2 m_1 c} \vec{\nabla}_1 \left( \vec{\nabla}_1. \vec{A}_1\right)+  \sum_j m_j \omega_j^2 \left(\vec{q}_j-\vec{r}_1-\vec{r}_2 \right)- k_1 (\vec{r}_1-\vec{r}_2)+ \notag \\
    &q_1 q_2\frac{\vec{r}_1-\vec{r}_2}{|\vec{r}_1-\vec{r}_2|^3}
    \label{Heisenbergpaticlemom1}
\end{align}
\begin{align}
    \dot{\vec{p}}_2=&\frac{1}{i\hbar}\left[\vec{p}_2,H  \right]=\frac{q_2}{c}\left( \vec{v}_2\times \vec{B} \right)+\frac{q_2}{c} \left(\vec{v}_2. \vec{\nabla}_1  \right)\vec{A}_2+\frac{i \hbar q_2}{2 m_2 c} \vec{\nabla}_2 \left( \vec{\nabla}_2. \vec{A}_2\right)+  \sum_j m_j \omega_j^2 \left(\vec{q}_j-\vec{r}_1-\vec{r}_2 \right)-k_2 \vec{r}_2 - \notag \\
    &q_1 q_2\frac{\vec{r}_1-\vec{r}_2}{|\vec{r}_1-\vec{r}_2|^3}
    \label{Heisenbergpaticlemom2}
\end{align}
\end{widetext}
   The derivatives of the vector potential $\vec{A}(r_k)$ (where $k = \{1,2\}$ represent the position coordinates $\vec{r}_1$ and $\vec{r}_2$ of the masses $m_1$ and $m_2$ respectively)
is derived using Eq.(\ref{mainHamiltonian}) as:\\
 
\begin{align}
    \dot{\vec{A}}(r_k)=\left(\vec{v}_k.\vec{\nabla}_k\right)\vec{A}(r_k)+\frac{i \hbar}{2 m_k}\nabla_k^2 \vec{A}(r_k) \label{Adot}
\end{align}
where $\vec{\nabla}_k=\frac{\partial}{\partial \vec{r}_k}$
Now, using Eq.(\ref{Adot}) and eliminating $\dot{\vec{p}}$ from Eqs.(\ref{Heisenbergpaticlemom1}),(\ref{Heisenbergpaticlemom2}), (\ref{Heisenbergpaticle1}) and (\ref{Heisenbergpaticle2}), we get:
\begin{align}
    m_1 \ddot{\vec{r}}_1=&\frac{q_1}{c}\left( \vec{v}_1 \times \vec{B} \right)+\sum_j m_j \omega_j^2 \left( \vec{q}_j-\vec{r}_1-\vec{r}_2\right)- \notag \\
   & k_1 (\vec{r}_1-\vec{r}_2) +q_1 q_2\frac{\vec{r}_1-\vec{r}_2}{|\vec{r}_1-\vec{r}_2|^3}\label{mrddot1}
\end{align}
\begin{align}
    m_2 \ddot{\vec{r}}_2=&\frac{q_2}{c}\left( \vec{v}_2 \times \vec{B} \right)+\sum_j m_j \omega_j^2 \left( \vec{q}_j-\vec{r}_1-\vec{r}_2\right)- \notag \\
    &k_2 \vec{r}_2 -q_1 q_2\frac{\vec{r}_1-\vec{r}_2}{|\vec{r}_1-\vec{r}_2|^3}\label{mrddot2}
\end{align}
In the next step, we derive the Heisenberg equations for the position and momentum coordinates of the bath oscillators:
\begin{align}
   \dot{\vec{q}}_{j}=\frac{1}{i\hbar}\left[\vec{q}_j,H\right] =\frac{\vec{p}_j}{m_j}
   \label{bathq}
\end{align}
\begin{align}
    \dot{\vec{p}}_{j}=\frac{1}{i\hbar}\left[\vec{p}_j,H\right]=-m_j \omega_j^2\left(\vec{q}_j-\vec{r}_1-\vec{r}_2 \right)\label{bathp}
\end{align}
Using Eqs.(\ref{bathq}) and (\ref{bathp}):
\begin{align}
\ddot{\vec{{q}}}_j+\omega_j^2 q_j=\omega_j^2\left(\vec{r}_1+\vec{r}_2  \right) \label{eombath}
\end{align}
Solving the Eq.(\ref{eombath}), we get the retarded solution for $\vec{q}_j$ as:
\begin{align}
    \vec{q}_j(t)=&\vec{q}_j^h (t)+\left(\vec{r}_1(t)+\vec{r}_2 (t) \right) \notag \\
    & -\int_{-\infty}^t dt' \cos{\left[\omega_j (t-t')\right]}\left(\dot{\vec{r}}_1(t')+\dot{\vec{r}}_2 (t') \right) \label{qjsol}
    \end{align}
where, $\vec{q}_j^h (t)$ is the solution of the homogeneous equation corresponding to Eq.(\ref{eombath})  and is a function of time $t$ and $\omega_{j}$. Later $\vec{q}_j^h (t)$ is used to eliminate the bath coordinate from particle variable solutions.\\
Replacing the bath coordinate $\vec{q}_j(t)$ in Eqs.(\ref{mrddot1}) and (\ref{mrddot2}) by Eq.(\ref{qjsol}), one can the equations of motion for the Brownian particles in a heat bath.
\begin{widetext}
\begin{align}
   & m_1 \ddot{\vec{r}}_1=\frac{q_1}{c}\left(\vec{v}_1 \times \vec{B} \right)-k_1\left(\vec{r}_1-\vec{r}_2 \right)+q_1 q_2\frac{\vec{r}_1-\vec{r}_2}{(r_1-r_2)^3}+ \sum_j m_j \omega_j^2 \vec{q}_j^h (t)- \int_{-\infty}^t  dt'\sum_j m_j \omega_j^2 \cos{\left[\omega_j\left(t-t' \right) \right]} \left[ \dot{\vec{r}}_1(t')+\dot{\vec{r}}_2(t') \right] \\
&  m_2 \ddot{\vec{r}}_2=\frac{q_2}{c}\left(\vec{v}_2 \times \vec{B} \right)-k_2 \vec{r}_2-q_1 q_2\frac{\vec{r}_1-\vec{r}_2}{(r_1-r_2)^3} +\sum_j m_j \omega_j^2 \vec{q}_j^h (t)- \int_{-\infty}^t  dt'\sum_j m_j \omega_j^2 \cos{\left[\omega_j\left(t-t' \right) \right]} \left[ \dot{\vec{r}}_1(t')+\dot{\vec{r}}_2(t') \right]
\end{align}
\end{widetext}
where we have used $  \vec{\nabla} \times \vec{B} =0$ as the current source of the external magnetic field lies outside the region of interest.\\
Therefore, the Langevin equation for the oscillators with coupled coordinates is given by:
\begin{align}
    & m_1 \ddot{\vec{r}}_1=\frac{q_1}{c}\left(\vec{v}_1 \times \vec{B} \right)-k_1\left(\vec{r}_1-\vec{r}_2 \right)+ \notag \\
     &q_1 q_2\frac{\vec{r}_1-\vec{r}_2}{(r_1-r_2)^3}+\vec{F}(t)  -\int_{-\infty}^t dt' \mu\left(t-t'\right) \left[\dot{\vec{r}}_1 (t')+\dot{\vec{r}}_2 (t')\right] \label{Langevinfinal1}
\end{align}
\begin{align}
     &m_2 \ddot{\vec{r}}_2=\frac{q_2}{c}\left(\vec{v}_2 \times \vec{B} \right)-k_2\vec{r}_2- \notag \\
     &q_1 q_2\frac{\vec{r}_1-\vec{r}_2}{(r_1-r_2)^3}+\vec{F}(t) -\int_{-\infty}^t dt' \mu\left(t-t'\right) \left[\dot{\vec{r}}_1 (t')+\dot{\vec{r}}_2 (t')\right]  \label{Langevinfinal2}
\end{align}
where,
\begin{align}
    \vec{F}(t)=\sum_j m_j \omega_j^2 \vec{q}_j^h (t)
\end{align}
\begin{align}
    \vec{\mu}(t)=\sum_j m_j \omega_j^2 \cos{(\omega_j t)} \Theta (t)
\end{align}
$\mu(t)$ and $\vec{F}(t)$ representing the memory kernel related to the damping parameter and the random force respectively \cite{bhattacharjee2022long}.\\
  As mentioned earlier \cite{ford1}, the multiple oscillator problem can be mapped into an IO (independent oscillator) scenario, considering the other oscillators as the heat bath particles. In a similar fashion, the case of coupled oscillators interacting with an Ohmic bath can mimic an IO model with the other particle behaving as an oscillator in the heat bath of infinite oscillators interacting with the system. Hence, the random forces satisfy the properties \cite{weiss2012quantum, Urbashione,surakaone}:\begin{align}
\langle{F_{\alpha}(t)}\rangle=0
\label{focecorr1}
\end{align}
\begin{align}
\notag \frac{1}{2}\langle{\lbrace F_{\alpha}(t),F_{\beta}(0)}\rbrace\rangle =\frac{\delta_{\alpha \beta}}{2 \pi}\int_{-\infty}^\infty{{d\omega}Re[\mu(\omega)]} \times \\
\coth\left(\frac{\hbar \omega}{2 K_B T}\right)
\hbar\omega e^{-i\omega t} \label{noisecor}
\end{align}
\begin{equation}
\langle{[F_{\alpha}(t),F_{\beta}(0)}]\rangle=\frac{\delta_{\alpha \beta}}{ \pi}\int_{-\infty}^\infty{{d\omega}Re[\mu(\omega)]}
\hbar\omega e^{-i\omega t}\label{noisecor2}
\end{equation}
Here $\alpha, \beta = x, y$ and $\delta_{\alpha \beta}$ is the Kronecker delta function.
Also 
$\mu(\omega)= \int_{-\infty}^{\infty}{dt \mu(t) e^{i\omega t}}$.\\
The Eq.(\ref{focecorr1}) represent the time average of the random forces in all directions and the Eqs.(\ref{noisecor}-\ref{noisecor2}) represent the commutation relations of the random noise which reduce to the $\delta$-correlated white noise at the high temperature Classical regime.\\
 It must be noted from Eqs.(\ref{Langevinfinal1}) and (\ref{Langevinfinal2}) that the dissipative term ($\mu$ dependent) for one particle is also affected by the  dissipation of the other particle caused by the interaction with the common heat bath. This similar kind of bath-mediated interaction was also seen in the case of two independent Brownian particles trapped close to each other and interacting with a common heat bath \cite{de2014nonequilibrium}. 
The results have significantly different outcomes corresponding to the dynamics of the individual masses in the coupled oscillator problem. \\
As pointed out earlier that the Coulomb repulsive term between the charges (Eqs. \ref{Langevinfinal1} and \ref{Langevinfinal2})
is inversely proportional to the square of the modulus
of separation between the charges. So the variation in
the Coulomb force is manifested only by the square of
the displacement of the particles $\Delta x_1 (\Delta y_1)$ and  $\Delta x_2 (\Delta y_2)$, which can be traced back from the Einstein’s relaxation mean square displacement: $\langle [\Delta x(t)]^2\rangle= 2 Dt$, where $D = K_B T / \zeta $ ($\zeta$ being the Stoke’s friction coefficient) \cite{chaikin_lubensky_1995}.
So at $T \rightarrow 0$, the variation in $\Delta x(t)$ is very small, even at long times.
Accordingly, we have considered the Coulomb repulsive
force to be constant in the following calculations for studying the effect of the magnetic field on the dynamics of the coupled oscillators. Here we solve the coupled Langevin equations (Eqs.(\ref{Langevinfinal1}) and (\ref{Langevinfinal2})) by taking the Fourier transform and breaking into the in-plane components:
\begin{equation}
    L \chi =\tilde{F} =F +F_c \label{Langevinomega}
\end{equation}
where, $\chi= (x_1, x_2, y_1, y_2)^{T}$, $\tilde{F}=(\tilde{F}_1, \tilde{F}_2, \tilde{F}_3, \tilde{F}_4)^{T}$ and $\tilde{F}_1=\tilde{F}_3=\tilde{F}_x$,  $\tilde{F}_2=\tilde{F}_4=\tilde{F}_y$
\begin{equation}
    L(\omega)=
    \begin{pmatrix}
     L_{11} & L_{12} &L_{13} &L_{14}\\
     L_{21} & L_{22} &L_{23} &L_{24}\\
     L_{31} & L_{32} &L_{33} &L_{34}\\
     L_{41} & L_{42} &L_{43} &L_{44}
    \end{pmatrix}
\end{equation}
where, $F_c$ represents the Coulomb repulsive force and the matrix coefficients: \\
$ L_{11} =L_{23}= -m_1 \omega^{2} + i \omega \mu(\omega) + k_1$, \\
$L_{12}= L_{24}=-k_1+i \omega \mu(\omega)$, \\
$L_{13}=-L_{21}=- i m_{1}\omega\omega_{c_1} $,\\
$L_{14}=L_{22}=L_{33}=L_{41}=0$; \\
$ L_{31} =L_{43}= -k_1+i \omega \mu(\omega)$ \\
$L_{32}= L_{44}=-m_{2}\omega^{2} + i\omega \mu(\omega)+k_2$\\ $L_{34}=-L_{42}=-i m_{2}\omega\omega_{c_2}$.

where $\omega_{c_i}=\frac{q_iB}{m_i c}$ is the cyclotron frequency of the $i^{th}$ particle and x$_1$, y$_1$, x$_2$, y$_2$ denotes the x and y coordinates for the first and the second particle respectively, as shown in the Fig.(\ref{schematic}). In the above Langevin equation (Eq.(\ref{Langevinomega})), we have considered $\omega_0=\sqrt{k_1/m_1}$ and  $\omega_0'=\sqrt{k_2/m_2}$. For simplicity, we have taken the cyclotron frequency to be the same for both the particles i.e., $q_i/m_i$=constant, such that $\omega_{c_1}=\omega_{c_2}=\omega_c$.

The position autocorrelation functions for the two particles $m_1$ and $m_2$ in the x-direction are respectively defined by:
\begin{align}
 C_{x_1}(t)&=\frac{1}{2}\langle \left\lbrace x_1(t),x_1(0) \right\rbrace \rangle \\
C_{x_2}(t)&=\frac{1}{2}\langle \left\lbrace x_2(t),x_2(0) \right\rbrace \rangle 
\end{align}
A similar relation holds for the y-axis too.\\
Following all the steps analogous to the case of a single oscillator \cite{bhattacharjee2022long}, one gets the following relation between the position correlation functions:
\begin{eqnarray}
C_{x_1}=\frac{1}{2}\langle \lbrace x_1(\omega),x_1^*(\omega)\rbrace \rangle=\frac{1}{2}\langle \lbrace y_1(\omega),y_1^*(\omega)\rbrace \rangle=C_{y_1}  \label{x1correlation}\notag \\ \\
C_{x_2}=\frac{1}{2}\langle \lbrace x_2(\omega),x_2^*(\omega)\rbrace \rangle=\frac{1}{2}\langle \lbrace y_2(\omega),y_2^*(\omega)\rbrace \rangle=C_{y_2}\notag \\\label{x2correlation}
\end{eqnarray}
The Response Function 
$R_x(\omega)$ is given by \cite{surakaone}:
\begin{equation}
    R_x (\omega)=Re \left[R_x(\omega)\right]+i Im \left[R_x(\omega)\right]
    \end{equation}
where $Im \left[R_x(\omega)\right]$ and $Re \left[R_x(\omega)\right]$ represent the imaginary part and the real part of the response function respectively
which can be directly obtained from the Quantum Langevin Equation
displayed in Eq.(\ref{Langevinomega}). Moreover, the imaginary part of this Response Function 
is connected to the spectral density (the Fourier transform of the position autocorrelation function) via the 
Fluctuation-Dissipation Theorem \cite{balescu}:
\begin{equation}
Im \left[R_x (\omega)\right]=R_x''(\omega)=\frac{1}{2\hbar}\left[1-\exp[-\beta \hbar \omega] \right]C_x(\omega)\label{FDT}
\end{equation}
So the position autocorrelation function ($C_{x}(t)$) can be written as:
\begin{align}
        &C_{x}(t)=\frac{1}{2\pi }\int_{-\infty}^{\infty} C_{x}(\omega) e^{-i \omega t}d\omega \\
       &  C_{x}(t)=\frac{\hbar}{\pi }\int_{-\infty}^{\infty} \frac{R_{x}''(\omega)}{\left[1-\exp[-\beta \hbar \omega] \right]}e^{-i \omega t}d\omega \label{Cxfouriertransform}
    \end{align}
   At T$\rightarrow$0, using Eq.(\ref{FDT}) one can get from Eq.(\ref{Cxfouriertransform}):
   \begin{align}
        &C_{x}(t)=\frac{\hbar}{\pi }\int_{0}^{\infty} \frac{R_{x}''(\omega)}{\left[1-\exp[-\beta \hbar \omega] \right]}e^{-i \omega t}d\omega \\
        &C_{x}(t)=\frac{1}{2\pi }\int_{0}^{\infty} C_{x}(\omega) e^{-i \omega t}d\omega 
   \end{align}
   
From the position autocorrelation function we obtain 
the position-velocity correlation function $C_{xv_x}(t)$ and 
the velocity autocorrelation function $C_{v_x}(t)$ by 
taking derivatives as follows:
\begin{equation}
C_{xv_x}(t)=\frac{d}{dt}C_x(t) \label{posvelcorr}
\end{equation}
and
\begin{equation}
 C_{v_x}(t)=-\frac{d^2}{dt^2}C_x(t)   \label{velautocorr}
\end{equation}
\section{Correlation functions of charged coupled oscillators in a magnetic field}
\subsubsection{Long-time tails in the position autocorrelation function}
\begin{widetext}

\begin{figure}[H]
\centering
\hspace*{-0.2cm}\includegraphics[scale=1.14]{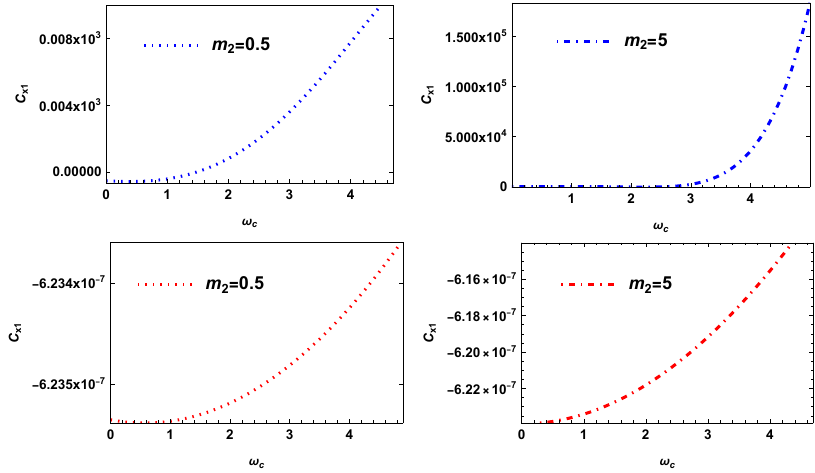} 
\caption{ $C_{x_1}$ vs. $\omega_c$ at different times for $\omega_0=0.1$, $\omega_0'=1$, $m_1=1$, $\mu=1$: \textbf{Blue} (top) and \textbf{Red} (bottom) curves represent the correlation functions at $\omega_0t=1$, $\omega_0t=10$ and $\omega_0t=100$ respectively.}
\label{f(magnetictimeplots)}
\end{figure}
\end{widetext}

In this sub-section, the long-time behaviour of the position autocorrelation function of two coupled  Brownian particles is studied in the presence of a harmonic oscillator potential using the framework used for the single particle case  \cite{bhattacharjee2022long}. Here, the calculation is done for the coupled oscillators attached to an Ohmic bath, with the bath particles executing harmonic oscillation \\
For an Ohmic bath, the damping parameter in Eq.(\ref{Langevinomega}) is given by  \cite{Urbashione,surakaone}:
\begin{equation}
    \mu_i(\omega)=m_i \gamma_i \label{Ohmickernel}
\end{equation}
where $\gamma$ is the dissipation constant.\\
In this paper, we have considered the bath-induced damping to be the same for both the particles: $\mu_1(\omega)=\mu_2(\omega)=\mu$.\\
Using, Eq.(\ref{Langevinomega})-(\ref{noisecor}) and Eq.(\ref{x1correlation})-(\ref{x2correlation}), one can obtain the power law behaviour of the position autocorrelation function for the particle with mass m$_1$ ($C_{x_{1}}(t)$):
\begin{align}
   &C_{x_{1}}(t) = \sum_{j=1}^\infty C_{x_{1}}^{(j)}(t)\\ 
  & C_{x_{1}}(t)=\sum_{j=1}^\infty A_j t^{-2j}
  \label{corrx1}
\end{align}
 where, 
 \begin{align}
 &A_1= -\hbar \mu \big[m_1^4 \omega_0^8 (m_1^2\omega_0^4-m_2^2\omega_0'^4)^2 + \notag \\
 &(m_2 \omega_0'^4 + m_2 \omega_0'^2 \omega_0^2 + 
     2 m_1 \omega_0^4)^2 \big] /  \big[2\pi m_1^6 \omega_0^{12}\times   \notag \\
     &(m_2\omega_0'^2 - m_1 \omega_0^2)^4\big] \label{A1}
 \end{align}
\begin{align}
    A_2&= A_2 (m_1,m_2,k_1,k_2,\omega_c) \notag\\
    &=A_{21}(m_1,m_2,k_1,k_2)+A_{22}(m_1,m_2,k_1,k_2) \omega_c^2+... \label{A2}
 \end{align}
Similarly, one can derive the position autocorrelation for particle with mass m$_2$ (C$_{x_2}$):
 \begin{align}
     C_{x_2}(t)=\sum_{j=1}^\infty B_j t^{-2j}\label{corrx2}
 \end{align}
 where,
 \begin{align}
  B_1= -\frac{4 \hbar \mu}{2\pi (m_2\omega_0'^2 - m_1 \omega_0^2)^2} \label{B1}
 \end{align}
 \begin{align}
     B_2&=B_2 (m_1,m_2,k_1,k_2,\omega_c) \notag\\
      &=B_{21}(m_1,m_2,k_1,k_2)+B_{22}(m_1,m_2,k_1,k_2) \omega_c^2+... \label{B2}
 \end{align}
It can be noted from Eqs.(\ref{A1})-(\ref{A2}) that the leading terms in the expansion of the correlation functions ($t^{-2}$) are independent of the magnetic field, whereas the sub-leading terms preserve the dominance of the magnetic field with the coefficients of $t^{-4}$ terms ($A_2$ and $B_2$) exhibiting a power law behaviour in terms of the cyclotron frequency $\omega_c$ (see Eqs.(\ref{A2})-(\ref{B2})). Hence the magnetic field plays an important role in determining the dynamics of the coupled oscillators at the moderately long time regimes by tuning the coefficient of the $t^{-4}$ terms. \\
 In Figs.(\ref{f(magnetictimeplots)}), we have plotted the position autocorrelation functions for the particle of mass $m_1$ as a function of the cyclotron frequency $\omega_c$ (dependent on the applied magnetic field) at different times for distinct values of the mass $m_2$. 

\begin{widetext}

\begin{figure}[H]
\centering
\includegraphics[scale=0.9]{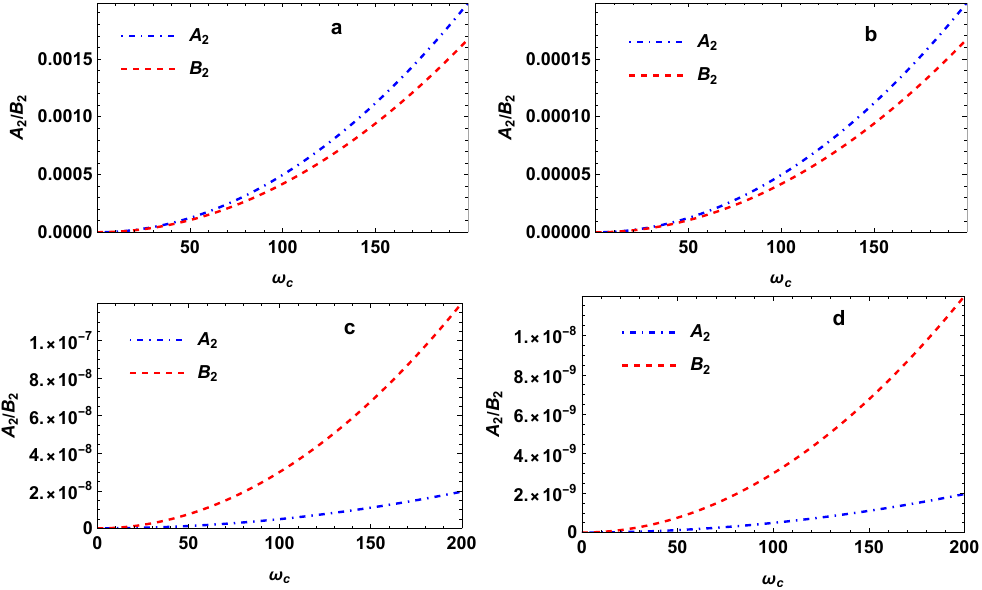} 

\caption{Variation of $A_2$ and $B_2$ with $\omega_c$ for $\omega_0$ = 0.1, $\omega_0'$=1, $\hbar$ = 1: (i) $m_1$ = 1, $m_2$ = 10 (a) Under-damped case -
$\mu$ = 20; (b) Over-damped case - $\mu$ = 2 and (ii) $m_1$ = 10, $m_2$ = 1 (c) Under-damped case - $\mu$ = 20; (d) Over-damped case - $\mu$ = 2}

\label{coefficient1}
\end{figure}
\end{widetext}

 The plots clearly show that the increase in magnetic field results in the increase of position autocorrelation function at an instant of time for the intermediate to long time regimes. Here, one can notice that the response to $\omega_c$ is more pronounced in the intermediate time regimes, whereas the correlation functions at the long time limit becomes almost independent of the magnetic field, due to the decay of the $\omega_c$-dependent sub-leading term in the expansion of $C_x$ (as seen from Fig(\ref{f(magnetictimeplots)})). As the correlation function in the intermediate time regime is primarily governed by the leading and the sub-leading terms in the series expansion, hence the coefficients of the sub-leading terms $A_2(B_2)$ play the significant role in determining the magnetic field-dependence of the correlation functions. So the variation of the coefficient $A_2$ with the masses $m_1$, $m_2$ and the damping parameter $\mu$ is manifested in the plots displayed in Fig.(\ref{coefficient1}).




 As discussed earlier, the magnetic field-dependent coefficients of the sub-leading order terms $A_2$ and $B_2$ exhibit an increasing trend with the rise in cyclotron frequency $\omega_c$ and the effect is further enhanced for lower values of the masses $m_1$ and $m_2$ (see Fig(\ref{coefficient1})), however, comparatively larger magnitude of $m_2$ imparts a more prominent $\omega_c$-dependence to the dynamics of the coupled oscillator system. This can be physically understood as 
 the oscillation due to the cyclotron frequency is more pronounced for the lighter masses, leading to the stronger magnetic field-effect on the coefficient corresponding to the masses $m_1$ ($A_2$) and $m_2$ ($B_2$). Moreover, it can be noted that the effect of $\omega_c$ is stronger in the under-damped regime, where the the magnetic field plays the dominant role over the damping coefficient $\mu$. \\
Now, from the above calculations (Eqs.(\ref{corrx1}) and (\ref{corrx2})), one can get the position correlation at zero magnetic field ($\omega_c=0$):
 \begin{align}
     C_{x_1}(t)(\omega_c=0)=C_{x_{1,0}}(t)=\sum_{j=1}^\infty A_{j,0}t^{-(j+1)}\\
      C_{x_2}(t)(\omega_c=0)=C_{x_{2,0}}(t)=\sum_{j=1}^\infty B_{j,0}t^{-(j+1)}
 \end{align}
 where,
 \begin{align}
 &A_{1,0}=A_1= -\hbar \mu \big[m_1^4 \omega_0^8 (m_1^2\omega_0^4-m_2^2\omega_0'^4)^2 + \notag \\
 &(m_2 \omega_0'^4 + m_2 \omega_0'^2 \omega_0^2 + 
     2 m_1 \omega_0^4)^2 \big] /  \big[2\pi m_1^6 \omega_0^{12}\times   \notag \\
     &(m_2\omega_0'^2 - m_1 \omega_0^2)^4\big] \label{A10}
\end{align}
 \begin{align}
   & A_{2,0}=A_{21}(m_1,m_2,k_1,k_2) \label{A20}
 \end{align}
 \begin{align}
 B_{1,0}=B_1=-\frac{4 \hbar \mu}{2\pi (m_2\omega_0'^2 - m_1 \omega_0^2)^2} \label{B10}
 \end{align}
  \begin{align}
      &B_{2,0}=B_{21}(m_1,m_2,k_1,k_2) \label{B20}
 \end{align}
\subsubsection{Long-time tails in the Position-Velocity Correlation Function}
The position-velocity correlation function for the two particles can be determined from the position autocorrelation function using Eq.(\ref{posvelcorr}):
\begin{equation}
    C_{xv_{x_k}}(t)=\frac{d}{dt}C_{x_k}(t)
=\sum_j C_{xv_{x_k}}^{(j)}(t)
\end{equation}

where
\begin{equation}
C_{xv_{x_k}}^{(j)}(t)=\frac{d}{dt}C_{x_k}^{(j)}(t)
\end{equation}
and $C_{x_k}$ and $C_{x_{k}}^{(j)}$ represent the x-coordinate of the position autocorrelation function for the $k^{th}$ mass and the $j^{th}$ term in it's power series expansion respectively. 


Hence, using Eqs.(\ref{corrx1}) and (\ref{corrx2}), we get,
\begin{align}
C_{xv_{x_1}}(t)=-\sum_j 2j A_j t^{-(2j+1)}
\end{align}
Similarly,
\begin{align}
C_{xv_{x_2}}(t)=-\sum_j 2j B_j t^{-(2j+1)}
\end{align}

Now, if the magnetic field is withdrawn, the coefficients $A_j$ and $B_j$ reduce to $A_{j,0}$ and $B_{j,0}$ respectively.
\subsubsection{Long-time tails in the Velocity Autocorrelation Function}
The velocity autocorrelation functions are determined from the position autocorrelation using Eq.(\ref{velautocorr})
\begin{equation}
C_{v_{x_{k}}}(t)=-\frac{d^2}{dt^2}C_{x_k}(t)=\sum_j C_{v_{x_k}}^{(j)}(t)
\end{equation}

where,
\begin{equation}
C_{v_{x_k}}^{(j)}(t)=-\frac{d^2}{dt^2}C_{x_k}^{(j)}(t)
\end{equation}

Then, using Eqs.(\ref{corrx1}) and (\ref{corrx2}), we get,
\begin{align}
C_{v_{x_1}}(t)=-\sum_j 2j(2j+1)A_j t^{-2(j+1)}
\end{align}
Similarly,
\begin{align}
C_{v_{x_2}}(t)=-\sum_j 2j(2j+1)B_j t^{-2(j+1)}
\end{align}

Similar to the case of position-velocity correlation, in the absence of magnetic field,  the coefficients of the power laws will be replaced by $A_{j,0}$ and $B_{j,0}$ respectively.\\
 In the Table-I, we have tabulated the power law behaviours (upto second order) of a single harmonic oscillator in a magnetic field \cite{bhattacharjee2022long} and the power law behaviours of the correlation functions for the two masses in a coupled harmonic oscillator system in the presence of a magnetic field and coupled to a common heat bath. It can be clearly seen from the table that the power laws remain same with the single oscillator case, however the coefficients vary distinctly, that is significant for the analysis of the coupled oscillator problem.

 \begin{widetext}
\begin{center}
\begin{table}[H]
 
\caption{ Table for the asymptotic expansion of the correlation functions}
\begin{tabular}{ |p{1.6cm}|p{2.5cm}|p{2.3cm}|p{2.4cm}|p{2.9cm}| p{2.5cm}|p{2.9cm}|}
\hline
 \multicolumn{3}{|c|}{Single Oscillator
 } & \multicolumn{2}{|c|}{Coupled Oscillator (Mass - $m_1$)
 } & \multicolumn{2}{|c|}{Coupled Oscillator (Mass - $m_2$)
 }\\
  \hline
  \hline
\centering{Parameters} &  \centering{$\omega_c=0$}  & \centering{$\omega_c \neq 0$} &  \centering{$\omega_c=0$} & \centering{$\omega_c \neq 0$} & \centering{$\omega_c=0$} & \centering{$\omega_c \neq 0$} \cr
\hline
 \centering{$C_x(t)$} &  \centering{$-\frac{\hbar }{\pi }P_1 t^{-2}+\frac{6\hbar}{\pi } P_{2,0}t^{-4}$}  & \centering{$-\frac{\hbar }{\pi }P_1 t^{-2}+\frac{6\hbar}{\pi } P_{2}t^{-4}$} &  \centering{$A_1 t^{-2}+A_{21} t^{-4}$} & \centering{$A_1 t^{-2}+\left[A_{21}+A_{22}  \omega_c^2\right] t^{-4}$} & \centering{$B_1 t^{-2}+B_{21} t^{-4}$} & \centering{$B_1 t^{-2}+\left[B_{21}+B_{22}  \omega_c^2\right] t^{-4}$} \cr
 \hline
 \centering{$C_{xv_x}(t)$} &  \centering{$\frac{2 \hbar}{\pi }P_1 t^{-3}-\frac{24 \hbar}{\pi }P_{2,0} t^{-5}$}  & \centering{$\frac{2 \hbar }{\pi  }P_1 t^{-3}-\frac{24 \hbar}{\pi } P_{2} t^{-5}$} &  \centering{$-2A_1 t^{-3}-4A_{21} t^{-5}$} & \centering{$-2A_1 t^{-3}-4 \left[A_{21}+A_{22} \omega_c^2 \right] t^{-5}$} & \centering{$-2B_1 t^{-3}-4B_{21} t^{-5}$} & \centering{$-2B_1 t^{-3}-4 \left[B_{21}+B_{22} \omega_c^2 \right] t^{-5}$} \cr
 \hline
  \centering{$C_{v_x}(t)$} &  \centering{$\frac{6 \hbar}{\pi }P_1 t^{-4}-\frac{120 \hbar}{\pi }P_{2,0} t^{-6}$}  & \centering{$\frac{6 \hbar }{\pi  }P_1t^{-4}-\frac{120 \hbar}{\pi } P_{2} t^{-6}$} &  \centering{$-6A_1 t^{-4}-20 A_{21} t^{-6}$} & \centering{$-6A_1 t^{-4}-20  \left[A_{21}+A_{22} \omega_c^2 \right] t^{-6}$} & \centering{$-6B_1 t^{-4}-20 B_{21} t^{-6}$} & \centering{$-6B_1 t^{-4}-20  \left[B_{21}+B_{22} \omega_c^2 \right] t^{-6}$} \cr
  \hline
\end{tabular}
\end{table}
\end{center}
 where, $P_1=\frac{\mu}{m\omega_0^4}$, $P_2=\frac{\mu(2 \omega_0^2+3 \omega_c^2-\mu^2)}{m \omega_0^8}$, $P_2(\omega_c=0)=P_{2,0}=\frac{\mu(2 \omega_0^2-\mu^2)}{m \omega_0^8}$, $A_1$, $A_{21}$, $A_{22}$, $B_1$, $B_{21}$, $B_{22}$ have the usual meanings as described in the text.
\end{widetext}

\section{Results and Discussions}
In this section, the position autocorrelation, position-velocity correlation and velocity autocorrelation function for the coupled harmonic oscillators are plotted for different damping regimes. The results are analyzed for two cases: (i) m$_1>$m$_2$ and (ii) m$_2>$m$_1$.
\begin{widetext}

\begin{figure}[H]

\centering

\includegraphics[scale=1.4]{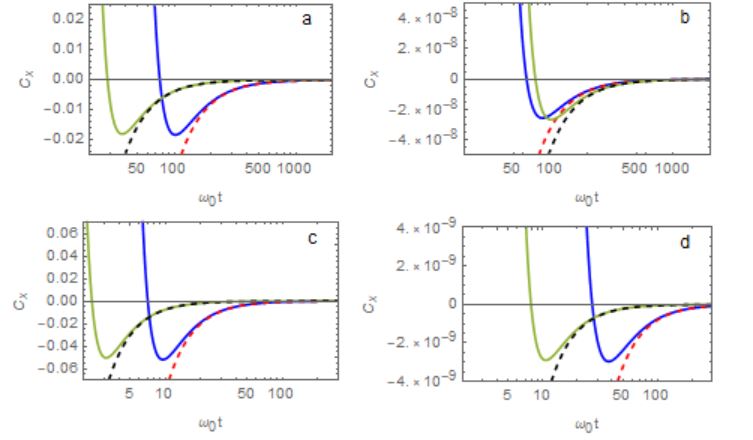} 
\caption{Position autocorrelation function versus time for  the over-damped ($\mu>\omega_c$) and under-damped ($\mu<\omega_c$) regimes with $\omega_c=1$: \textbf{(i)} m$_2>$m$_1$ ($m_1=1$, $m_2=10$), $\omega_0=0.1$, $\omega_0'$=1, $\hbar=1$: (a) Over-damped- $\mu=20$; (b) Under-damped regions- $\mu=2$ and \textbf{(ii)} m$_1>$m$_2$ ($m_1=10$, $m_2=1$), $\omega_0=0.1$, $\omega_0'$=1, $\hbar=1$: (c) Over-damped-  $\mu=20$; (d) Under-damped-  $\mu=2$ regions. The blue and the green curves represent C$_{x_1}$ and C$_{x_2}$ respectively. The red dashed curve and the black dashed curve represent the power law behaviour of C$_{x_1}$  and C$_{x_2}$ respectively. 
\vspace{8cm}} 
\label{posautocorr}
\end{figure}
\end{widetext}

\begin{widetext}

\begin{figure}[H]
\centering
\includegraphics[scale=1.3]{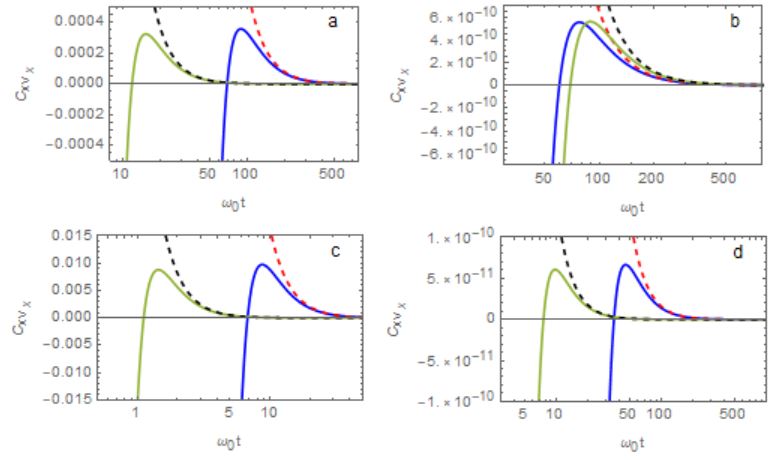} 
\caption{Position-velocity correlation function versus time for  the over-damped ($\mu>\omega_c$) and under-damped ($\mu<\omega_c$) regimes with $\omega_c=1$: \textbf{(i)} m$_2>$m$_1$ ($m_1=1$, $m_2=10$), $\omega_0=0.1$, $\omega_0'$=1, $\hbar=1$: (a) Over-damped- $\mu=20$; (b) Under-damped  regions- $\mu=2$ and \textbf{(ii)} m$_1>$m$_2$ ($m_1=10$, $m_2=1$), $\omega_0=0.1$, $\omega_0'$=1, $\hbar=1$: (c) Over-damped-  $\mu=20$; (d) Under-damped-  $\mu=2$ regions. The blue and the green curves represent C$_{x_1}$ and C$_{x_2}$ respectively. The red dashed curve and the black dashed curve represent the power law behaviour of C$_{x_1}$  and C$_{x_2}$ respectively.} 
\label{posvelcorr}
\end{figure}

\end{widetext}

\begin{widetext}
    
\begin{figure}[H]
\centering
\includegraphics[scale=1.3]{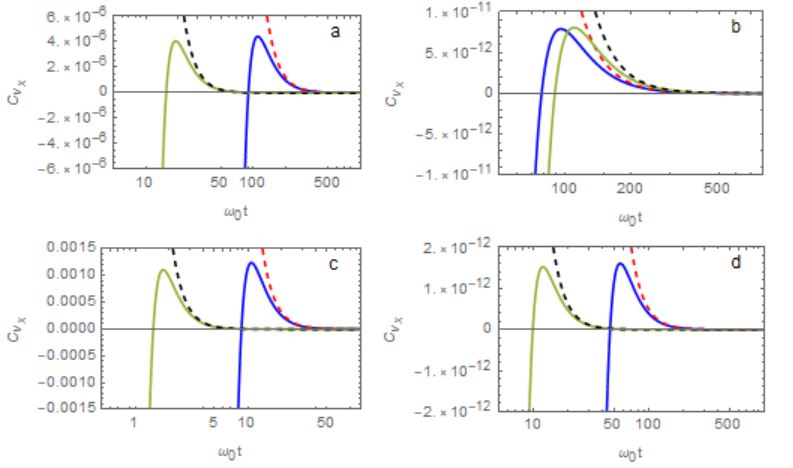} 
\caption{Velocity autocorrelation function versus time for   the over-damped ($\mu>\omega_c$) and under-damped ($\mu<\omega_c$) regimes with $\omega_c=1$: \textbf{(i)} m$_2>$m$_1$ ($m_1=1$, $m_2=10$), $\omega_0=0.1$, $\omega_0'$=1, $\hbar=1$: (a) Over-damped- $\mu=20$; (b) Under-damped regions- $\mu=2$ and \textbf{(ii)} m$_1>$m$_2$ ($m_1=10$, $m_2=1$), $\omega_0=0.1$, $\omega_0'$=1, $\hbar=1$: (c) Over-damped-  $\mu=20$; (d) Under-damped-  $\mu=2$ regions. The blue and the green curves represent C$_{x_1}$ and C$_{x_2}$ respectively. The red dashed curve and the black dashed curve represent the power law behaviour of C$_{x_1}$  and C$_{x_2}$ respectively.}
\label{velautocorr}
\end{figure}
\end{widetext}
The plots in Figs.[\ref{posautocorr}-\ref{velautocorr}] show that the correlation functions initially show an oscillatory behaviour due to the cyclotron frequency and then decay as time increases.  The rate of damping is more pronounced for the over-damped cases compared to the under-damped region, which is evident from the time scale presented in the plots. The damping exhibits the $t^{-2}$ power-law decay for position autocorrelation, $t^{-3}$ decay for position-velocity correlation and $t^{-4}$ behaviour for velocity autocorrelation functions at long times, as shown in the previous section.  It can be further noticed that the plots of the under-damped cases follow the power law behaviour of the leading terms at longer times (Figs.(\ref{posautocorr}b),(\ref{posvelcorr}b), (\ref{velautocorr}b)) with respect to their over-damped (Figs.(\ref{posautocorr}a),(\ref{posvelcorr}a), (\ref{velautocorr}a)) counterparts. This outcome is physically expected from the analytical results as the sub-leading terms contain the effect of cyclotron frequency and the effect of magnetic field is more prominent in the under-damped cases, where the damping occurs at a relatively slower rate. So the sub-leading term is expected to retain its importance up to a comparatively longer time scale for the under-damped region, as can be seen from Figs.[\ref{posautocorr}-\ref{velautocorr}].  Moreover, one can notice that the effect of increased damping ($\mu$) is clearly visible in case of the correlation functions of the particle of mass m$_2$ (C$_{x_2}$, C$_{xv_{x_2}}$, C$_{v_{x_2}}$). This can be understood from the analytical calculation that we have already mentioned. Furthermore, the effect of the oscillatory behaviour due to the magnetic field is also more dominant resulting in the slower decay of the correlation function in the under-damped regime. \\
\hspace*{0.2cm}The calculation of the correlation functions and the power law behaviours of a particle oscillating harmonically in a magnetic field was studied in the framework of the Langevin formalism \cite{bhattacharjee2022long}. However, in this work, we have revisited a similar kind of approach and derived the correlation functions for two coupled harmonic oscillators in a magnetic field at T$\rightarrow 0$. The coupling of the two oscillators preserves the power law  behaviours for a single particle executing quantum Brownian motion \cite{bhattacharjee2022long}, but the coefficients of the power laws are drastically changed, affecting the overall nature of the fall of correlation functions. The effect of the magnetic field is thoroughly analyzed in the over-damped and under-damped regimes and visibly distinct results are obtained for the two particles, which have different consequences from the physical point of view. Although, some earlier works have focused on the coupled classical Brownian motion particles \cite{Berkowitz, ford2014}, but, the results for the correlation functions of charged coupled harmonic oscillators in an external magnetic field and interacting with an Ohmic heat bath are entirely new in the backdrop of Quantum Brownian motion. The outcomes in the absence of the magnetic field are also presented in the context of the motion of two neutral coupled oscillators in the presence of a common external heat bath \cite{de2014nonequilibrium}.
\section{Conclusion}
In this paper, we have revisited the Langevin dynamics of two coupled oscillators in presence of an external magnetic field and connected to an Ohmic heat bath. We have shown that in the common heat bath problem, the interaction between the bath and one oscillator indeed affects the dynamics of the other oscillator sharing the common bath,  in agreement with the study of two Brownian particles trapped close to each other by an external potential \cite{de2014nonequilibrium}. Furthermore, the analysis of two coupled oscillators connected by a string exhibits that this case can be mapped into an IO (independent oscillator) scenario only when the length of the string $L\rightarrow\infty$ \cite{ford2014}. However, in our analysis we have shown a similar kind of bath-mediated interactions between the particles, originating from the oscillators in the common Ohmic heat bath, using an IO formulation for the force correlation functions. This is analogous to the Casimir-Polder effect and Casimir effect, which explains the interaction between two parallel uncharged plates in vacuum, due the fluctuation of the electromagnetic field and the interaction is mediated by the quantized photons\cite{casimir,Panella}. In the macroscopic scenario, the effect corresponds to the Casimir effect, however the soft-photon renormalizations has to be taken into account for the charged particles \cite{Panella}.  Thus the coupled oscillator scenario is significantly different from the single oscillator case and can not be treated as a mere extension of the single oscillator problem.\\
\hspace*{0.2cm}The position autocorrelation, position-velocity correlation and velocity autocorrelation functions of the two masses are separately studied in the presence of a magnetic field and for T$\rightarrow$0. The moderately long-time behaviour of the correlation functions is derived, where the sub-leading terms in the series expansion are expected to play a significant role together with the leading term. The effect of the magnetic field on the coefficient of the sub-leading order term is analytically and numerically analyzed in the different damping regions and for distinct values of the spring constant $k_1$ and $k_2$. \\
\hspace*{0.2cm} The analytical results and the plots show that the power law trends remain the same as the single particle oscillator, however, there is a drastic change in the coefficients of the power laws that affect the over trends of decay (see Figs. (\ref{posautocorr}-\ref{velautocorr})). Moreover, the correlation functions for the particles m$_1$ and m$_2$ are distinct and are individually governed by the single oscillator forces and the extra force exerted due to the presence of the second particle and mediated by the common heat bath.  \\
\hspace*{0.2cm} In \cite{Berkowitz,ford1}, the Langevin dynamics for coupled oscillators executing Brownian motion was classically studied. Furthermore, in one of our previous papers, we had stressed on the long time behaviours of correlation functions of a charged quantum particle in the presence of a magnetic field \cite{bhattacharjee2022long}. However, here we have gone beyond these results and derived the long time behaviours of the position autocorrelation, position-velocity correlation and velocity autocorrelation of two coupled charged oscillators, executing Quantum Brownian motion and highlighted the effect of the magnetic field on the dynamics of the coupled oscillator. \\
\hspace*{0.2cm}  The outcomes have importance in analysing the dynamics of two coupled pendulums in connection to a heat bath \cite{Nakagawa}.  In addition, these results can be extremely important for the experimental realization of Quantum Brownian motion of molecules possessing more than one atom and the oscillations of one atom is affected by the stretching of bonds of the neighbouring atoms. As for example, the results presented above in the absence of a magnetic field are applicable for studying the small atoms in a protein molecule at very low temperature, where the Quantum fluctuations play the dominating role over the thermal fluctuations \cite{McCammon,McCammon1}.  Further, the results can be experimentally verified in cold atom laboratories by trapping the atoms using optical or magneto-optical traps.  \\
\section{Acknowledgement}
The authors acknowledge Prof$.$ Supurna Sinha for the discussions regarding the conceptualization and formulation of the problem.\\
\end{document}